\documentclass[doublecol,figures]{manuscript} 
\usepackage{amssymb}

\title{Quantum turbulence at finite temperature: \\
the two--fluids cascade}
\author{P.-E. Roche{$^1$},   C.F. Barenghi{$^2$}  \and E. Leveque{$^3$}  }
\institute{
\inst{1} Institut N\'eel, CNRS/UJF,
 BP166, F-38042 Grenoble Cedex 9, France\\
 \inst{2} School of Mathematics and Statistics, Newcastle University,
 Newcastle upon Tyne NE1 7RU, UK\\
\inst{3}  Laboratoire de Physique, ENS Lyon, CNRS/Universit\'e de Lyon
F-69364 Lyon, France
 }

\pacs{47.37.+q}{Hydrodynamic aspects of superfluidity: quantum fluids}
\pacs{47.27.ek}{Direct numerical simulations}
\pacs{47.27.Gs}{Isotropic turbulence; homogeneous turbulence}
\abstract{
To model isotropic homogeneous quantum turbulence in superfluid helium, we have performed Direct Numerical Simulations (DNS) of two fluids (the normal fluid and the superfluid) coupled by mutual friction. We have found evidence of strong locking of superfluid and normal fluid along the turbulent cascade, from the large scale structures where only one fluid is forced down to the vorticity structures at small scales. We have determined the residual slip velocity between the two fluids, and, for each fluid, the relative balance of  inertial, viscous and friction forces along the scales. Our calculations show that the classical relation between energy injection and dissipation  scale is not valid in quantum turbulence, but we have been able to derive a temperature--dependent superfluid analogous relation. Finally, we discuss our DNS results in terms of the current understanding of quantum turbulence, including the value of the effective kinematic viscosity.
}

\newcommand{\vn}{\mathbf {v}_n}
\newcommand{\vs}{\mathbf {v}_s}
\newcommand{\vns}{\mathbf {v}_{ns}}

\newcommand{\fns}{\mathbf {F}_{ns}}
\newcommand\bom{{\mbox{\boldmath $\omega_s$}}}
\newcommand\bomn{{\mbox{\boldmath $\omega_n$}}}

\newcommand{\ron}{\rho_n}

\newcommand{\ros}{\rho_s}

\newcommand{\pn}{p_n}
\newcommand{\ps}{p_s}

\begin{document}

\maketitle

\section{Motivation and aim}

The low temperature phase of liquid helium $^4$He (He~II) consists of two
co--penetrating fluids\cite{Donnelly}: 
an inviscid superfluid (associated to the quantum
ground state) and a gas of thermal excitations which make up the
viscous normal fluid. Quantum mechanics
constrains the rotational motion of the superfluid to discrete,
quantised vortex filaments of fixed circulation $\kappa$; these vortices
scatter the thermal excitations, thus inducing a mutual friction force
between the two fluids\cite{BarenghiJLTP1983}. 
Mechanical \cite{Maurer1998,Smith1999,Stalp2002,RocheEPL2007} or
thermal driving\cite{Vinen:1958ys} can easily excite turbulence
in both fluids.  The resulting state of quantum turbulence
at temperatures \footnote{In the zero-temperature limit, the
normal fluid, hence the mutual friction, is negligible, and the
quantum turbulence problem becomes different}
above $1~\rm K$
and its similarities with ordinary turbulence is a problem which is
attracting interest, and is
the subject of this investigation.

According to experimental, theoretical and numerical results 
(for a review see \cite{Vinen2002}),
at sufficiently large scales in the inertial  range, the superfluid and
normal fluid velocities are matched ($\vs \approx \vn$) and their spectra
obey the classical  Kolmogorov scaling $k^{-5/3}$ (where $k$ is the magnitude
of the three--dimensional wavenumber).
Our aim is to go beyond this first order description and explore
the consequences of the finiteness of the mutual coupling between the two
fluids in the inertial and dissipative ranges of the
turbulent cascade. In particular we want to know if the inter--fluid 
locking remains efficient at different
temperature, 
what is the residual slip velocity between the two
fluids and
how the energy transfer between the two fluids affects the
classical formulae from ordinary turbulence theory which
relate injection and dissipation to  Reynolds number.
%XXX what is the spectrum of the superfluid vorticity (cela ne rentre pas dans la categorie finite coupling
To answer these questions we introduce a numerical model based on 
the traditional direct numerical simulation (DNS) of the
Navier--Stokes equation.

%TO REVISE
%Putting forward a few theoretical arguments, we argue that several 
%features evidenced by the simulations should also be present 
%in turbulent He-II. We first generalize recent numerical 
%results on the 2 fluids locking when mutual coupling 
%is taken into account both ways. We then show that the residual 
%slip between the 2 fluids in the inertial range can be determined 
%by writing that it allows energy transfer between the fluid to 
%balance exactly the viscous dissipation. This results in a
% peaking of the slip-velocity at small scales. Finally, we argue 
%that the energy transfer between the 2 fluids near the 
%dissipative cut-off scales is responsible for a temperature-dependence 
%of the flow Re number, defined using the ratio of the largest and 
%smallest scales of turbulence, for given kinematic viscosity 
%and total energy injection. To understand quantitatively this 
%observation, we derive a superfluid effective viscosity, which is 
%found consistent with the microscopic derivation of Vinen \cite{Vinen2002} 
%and reproduces over nearly 2-decade the variation of (another) 
%effective viscosity reported in experiments (\cite{Walmsley:PRL2008} 
%and ref. within). A side result is the suggestion that the
% separation between the regimes of quantum turbulence at zero-temperature 
% and at finite-temperature is near 500 mK and not 1K as usually considered.

\section{Numerical model}

Our model is inspired by the HVBK equations 
\cite{Hall:1956ly,Bekarevich}, which 
describe the dynamics of He~II in the continuum limit 
using a Navier--Stokes equation (for the normal fluid) and an 
Euler equation (for the superfluid) coupled by a mutual friction 
force. The HVBK equations have been used with success to describe
vortex waves in rotating helium\cite{Henderson2004} and to predict the
observed instability of the Taylor--Couette flow of
He~II between two rotating concentric cylinders in the linear
\cite{Barenghi1988,Barenghi1992} and nonlinear regimes\cite{Henderson1995}.
The key aspect of the HBVK equations is that they smooth out the 
discrete nature of superfluid vortex 
lines by introducing a ``coarse--grained'' superfluid vorticity field $\bom$. 
The advantage of the HVBK approach is that it allows us to
account for the fluid motion on scales larger than the typical 
inter--vortex spacing $\ell$
in a dynamically self--consistent way, that is to say, the normal fluid 
determines the superfluid and vice versa. 
The disadvantage in the context of turbulence is that 
vortex filaments which are randomly oriented and  Kelvin waves of
wavelength smaller than $\ell$ do not contribute to $\bom$;
hence, if we define the vortex line density (vortex length per unit volume)
as $L=\vert \bom \vert /\kappa$, we underestimate its actual value.
In other words, the HVBK approach only captures 
the ``polarised'' contribution of the superfluid vortex tangle\cite{Roche2008}.

For reference, we note that a numerical simulation of
quantum turbulence with friction coupling acting (self--consistently)
on both fluids has already been reported using Large Eddy Simulation 
\cite{Merahi:2006}.  The self--consistent coupling of Schwarz's
vortex filament method\cite{Schwarz1988}
 for the superfluid (which models the motion
of individual vortex lines) and a Navier--Stokes equation
for the normal fluid has been implemented only for the case of a single
vortex ring \cite{Kivotides-Barenghi-Samuels-2000}. Modifications
of Schwarz's method which are not dynamically self--consistent (in which
the normal fluid affects the superfluid vortices but not vice versa) have
also been proposed 
\cite{Barenghi1997,Kivotides2002,Kivotides2007,Morris:PRL2008}.

The model which we propose differs from the HVBK equations
in two respects. Firstly, we introduce an artificial superfluid 
viscosity $\nu_s$ as a ``closure'' to damp the energy at the smallest 
scales and prevent possible numerical instabilities and numerical artifacts; 
this feature also greatly simplifies our task, as it allows us to make use of efficient Navier--Stokes validated codes, optimised to run on parallel supercomputers.
 The value of $\nu_s$
is chosen to be as small as possible, smaller than the normal
fluid's value $\nu_n$, so that the artificial damping of the superfluid
occurs at smaller scales than viscous damping of the normal fluid. 
Secondly, we simplify the form of the mutual friction force of the HVBK 
equations, to allow a more direct interpretation. 
The resulting equations for the normal fluid and superfluid velocity 
fields $\vs$ and $\vn$ are
\begin{eqnarray}
\label{eq:main}
\frac{D \vn}{D t} = -\frac{1}{\ron} \nabla \pn  + \frac{\rho_s}{\rho} \fns 
                    + \nu_n \nabla^2 \vn + {\bf f}_n^{ext},\\
\frac{D \vs}{D t} = -\frac{1}{\ros} \nabla \ps   - \frac{\rho_n}{\rho}\fns
                    + \nu_s \nabla^2 \vs + {\bf f}_s^{ext},
\end{eqnarray}

\noindent
where the indices $n$ and $s$ refer to the normal fluid and superfluid 
respectively, ${\bf f}_{n}^{ext}$ and ${\bf f}_{s}^{ext}$ are 
external forcing terms, $\rho_{n}$ and $\rho_{s}$ are the normal fluid
and superfluid densities, $\rho=\rho_n+\rho_s$, $\pn=(\rho_n/\rho)p+\rho_s ST$ and 
$\ps=(\rho_s/\rho) p -\rho_s ST$ are partial pressures, $S$, $T$ and $p$ are
specific entropy, temperature and pressure, and
$\vn$ and $\vs$ satisfy the incompressibility conditions
$\nabla \cdot \vn=0$ and $\nabla \cdot \vs=0$.
For simplicity the mutual friction force is written as
\begin{equation}
\fns=\frac{B}{2} \vert \bom \vert \vns,
\label{eq:fns}
\end{equation}

\noindent
where $\vns=\vn-\vs$ is the slip velocity and
$\bom=\nabla \times \vs$ is the superfluid vorticity.

Our numerical code for two--fluids DNS has been adapted from an existing single--fluid DNS code (used for example in \cite{Leveque:2001}).  Here it suffices to say 
that it is based on a pseudo--spectral method with
$2^{\rm nd}$ order accurate Adams--Bashforth time stepping; the computational
box is cubic (size $2\pi$) with periodic boundary conditions in the three directions and the spatial
resolution is $256^3$. Validation of the code includes checking the
preservation of the solenoidal conditions and of the correct balances of 
the energy fluxes (injected, exchanged between the two fluids and dissipated).

Calculations were performed with density ratios $\rho_n/\rho_s$  
$10$, $1$ and $0.1$, corresponding respectively to 
$T=2.157~K$ (hereafter referred to ``high temperature''), $T=1.96~K$ 
(``intermediate temperature'') and $T=1.44~K$ (``low temperature'')
\cite{DonnellyBarenghi1998}
In this range of temperatures the friction coefficient $B$ changes
only by a factor of two, thus, to facilitate the interpretation of the
results, we set $B=2$ constant in all calculations.
At each temperature, we set the artificial kinematic viscosity of 
the superfluid to be four time smaller than the normal fluid kinematic 
viscosity, $\nu_n/\nu_s=4$. One extra calculation, performed at 
high temperature with $\nu_n/\nu_s =100$, will be hereafter
referred to as the ``Quasi-Euler'' superfluid case, because energy 
dissipation by the superfluid viscosity is 
negligible\footnote{more than 99.8\% of the injected kinetic 
energy was dissipated by the normal fluid}. 
A random forcing 
(acting in the shell of wave-vectors $1.5 < |\mathbf{k}| < 2.5$) 
was imposed on the normal fluid alone at high and 
intermediate temperatures, and on the superfluid alone at low temperature. 
The intensity of this forcing was such that the total energy injected 
per unit mass of fluid was constant over time, and was kept the same 
at all temperatures.

\section{Results}
\subsection{The locking of the two fluids}

\begin{figure*} 
\begin{center} 
\includegraphics[scale=.25]{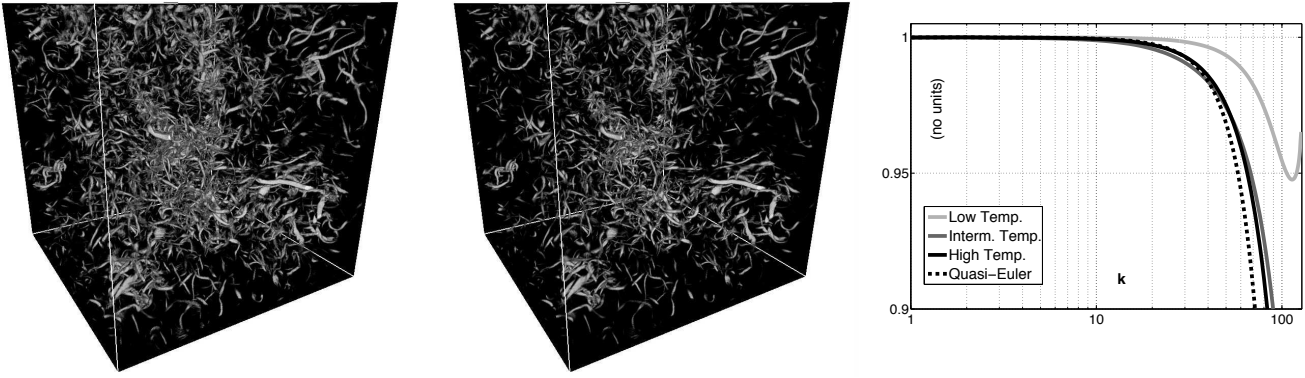} 
\caption{Left and Center visualisations : local enstrophy (squared vorticity) field  
of superfluid (left) and normal fluid (center) at intermediate temperature. 
The pattern of vorticity structures is the same
in both fluids, but is more intense in the superfluid.
The grayscale is defined in the text. Right plot : Spectral coherence function $c_2(k)$ of 
normal and superfluid velocity fields.} 
\label{fig:Visu} 
\end{center} 
\end{figure*} 

Figure \ref{fig:Visu} shows the local superfluid and normal fluid enstrophies 
$\vert \bom \vert^2$ (left) and  
$\vert \bomn \vert ^2$ (right) at the intermediate temperature 
($\rho_n=\rho_s$). Intense vortex regions (``worms'') are highlighted by 
the colour scale,
which becomes gray above $5 <\vert \bom \vert ^2>$ and 
opaque white above $10 <\vert \bom \vert ^2>$, 
where the symbol $<\cdots>$ indicates volume averaging
\footnote{This visualisation was generated with ``Vapor'' freeware, 
downloadable at www.vapor.ucar.edu}.
The figure clearly shows that the vorticity structures of the
two fluids are very similar. The degree of similarity can be quantified 
by the correlation coefficient 
\begin{equation}
c_1 = \frac{<\bom \bomn >}{ \sqrt{ <\bom ^2><\bomn ^2>} },
\label{eq:c1}
\end{equation}

We find that $c_1$ is larger than $97\%$ at all three temperatures explored,
in agreement with recent numerical calculations \cite{Morris:PRL2008} 
of a superfluid vortex tangle driven by a 
turbulent normal fluid at high temperature ($\rho_n/\rho_s \simeq 3$). 
Unlike our work, these simulations ignored the back--reaction of the 
vortex tangle onto the normal fluid. 

%\begin{figure}[!ht]
%\includegraphics[width=\linewidth]{CoherenceCinesNB.eps}
%%\includegraphics[scale=0.4]{fig2.pdf}
%\caption{Spectral coherence function $c_2(k)$ of 
%normal and superfluid velocity fields.}
%\label{fig:Coherence}   
%\end{figure}

The plot on the right side of Fig.~\ref{fig:Visu} compares the spectral coherence function
\begin{equation}
c_2(k) = \frac{| \vn(k) . \overline{\vs(k) }|^2}{| \vn(k) |^2 . | \vs(k) |^2 }
\label{eq:c2}
\end{equation}

\noindent
of superfluid and normal fluid velocity fields:
The fact that $c_2(k)$ is larger than 98\% at all scales below 
$k_{max}/4$, where $k_{max}=128$ is the largest wave vector associated with our 256$^3$ mesh, 
means that superfluid and normal fluid velocity
fields are strongly locked along the inertial range, up to the 
forcing length scale, where the energy, which is injected into one fluid
only, is redistributed very efficiently over both fluids. We recall that energy is injected into the normal fluid only,
except at low temperature, where it is injected in the superfluid only.

\begin{figure}[!ht]
\includegraphics[width=\linewidth]{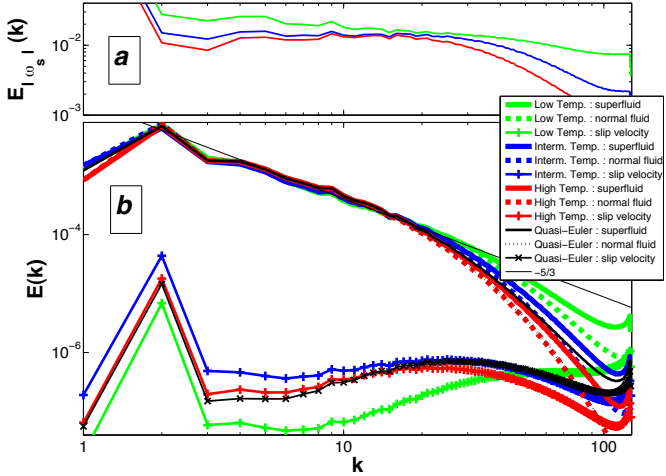}
\caption{[colour online]
Subplot a : Spectra of the magnitude of the superfluid vorticity $\vert \bom \vert$ vs wavenumber $k$. From top to bottom low [green], intermediate [blue] and high [red] temperatures.
Subplot b / Upper part:
Superfluid (thick solid lines) and normal fluid (thick dashed lines)
velocity power spectra vs wavenumber $k$
at the three temperatures. The thin dashed line denotes the
Kolmogorov $k^{-5/3}$ scaling. Subplot b / Lower part:
The thin solid lines show the spectra of the slip velocity 
at the three temperatures.}
\label{fig:Spectre}   
\end{figure}

The upper set of curves of Fig.~\ref{fig:Spectre}b presents the
superfluid and normal velocity power spectra $E_n(k)$ and $E_s(k)$
(defined by $E_i=\int E_i(k) dk$ ($i=n,s)$ where $E_i$ is the 
total energy in arbitrary units)
at high, intermediate and low temperature.
As expected from the previous figure, the spectra overlap along the 
inertial range; around $k=10$, both spectra are compatible with 
Kolmogorov's $k^{-5/3}$ inertial range scaling (illustrated by the
thin straight line). Note also that at
each temperature, since $\nu_s<\nu_n$, 
for increasing $k$ the normal fluid becomes damped before the superfluid.
This is consistent with Fig. \ref{fig:Visu}, which shows that
the superfluid enstrophy is indeed stronger than the normal fluid's. 

\subsection{The slip between the two fluids}

If the two fluids were perfectly
locked, the mutual friction would be zero, which would inevitably result in the
unlocking of the two fluids, as they would experience different forcing and 
dissipation processes. A residual slip velocity $\vns=\vs -\vn$
between the two fluid must therefore be present. 
The lower set of curves in Fig.\ref{fig:Spectre}b shows the spectrum of 
$\vns$. The peak at $k\approx 2$ is caused by the forcing, 
which is applied to a single fluid, and induces a residual slip 
at this wavevector. The striking feature of this spectrum 
is that it increases with $k$ in the inertial range, starting
at $k \sim 10$, and peaking in the dissipation scales, where the dissipation
of one fluid is significantly larger than that of the other.
The increase with $k$ of the slip velocity spectrum 
is remarkably pronounced in the ``quasi-Euler'' case.

\subsection{Energy transfer between the two fluids}

We focus on the high temperature ``quasi-Euler'' case 
($\nu_n/ \nu_s=100$, $\rho_n/ \rho_s=10$) because it is
expected to mimic He~II hydrodynamics more closely than other
viscosity ratios. 
The scale--by--scale energy budget per mass unit for the normal fluid 
and superfluid are respectively:

\begin{eqnarray}
\label{eq-budget}
\frac{\partial E_n}{\partial t}(k,t) = - D_n(k,t) - T_n(k,t) 
          -M_{n \to s}  + \epsilon^{inj} \delta _{k,2},\\
\frac{\partial E_s}{\partial t}(k,t) = - D_s(k,t) - T_s(k,t) 
          -M_{s \to n},  
\end{eqnarray}

\noindent
where $D_{n}$ and $D_{s}$ are the viscous dissipation terms in 
the normal fluid and superfluid,
\begin{equation}
\label{eq-D}
D_{i}  (k,t) = 2 \nu_{i} k^2 E_{i}(k,t)
\end{equation}

\noindent
$T_{n}$ and $T_{s}$ are the energy transfer rates arising from
triad interactions between Fourier modes within each fluid; 
the energy flux at wave number k related to the non-linear interaction 
is defined by
\begin{equation}
\label{eq-F}
F_{i} (k,t) = \int_{0}^k T_{i}(k',t) dk'.
\end{equation}

\noindent
$M_{s \to n}$ and $M_{n \to s}$ result from the exchange of kinetic energy 
between the two fluids by mutual friction :

\begin{equation}
\label{eq-Csn}
M_{n \to s} (k,t) = - \frac{\rho_{s}}{\rho} (\fns \cdot \vn) (k,t),
\end{equation}

\begin{equation}
\label{eq-Cns}
M_{s \to n} (k,t) =  \frac{\rho_{n}}{\rho} (\fns \cdot \vs) (k,t).
\end{equation}

\begin{figure}[!ht]
\includegraphics[width=\linewidth]{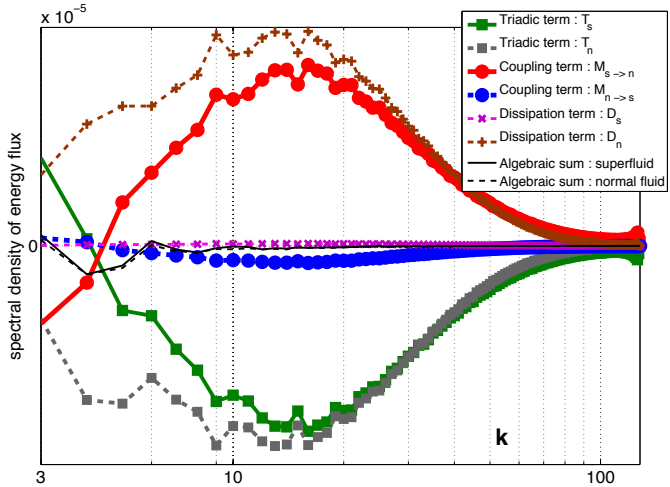}
\caption{[colour online] Energy exchange between fluids in the high temperature ``quasi-Euler'' case ( $\rho_{n}/\rho_{s}=10$ and $\nu_{n}/\nu_{s}=100$).}
\label{fig:Flux}   
\end{figure}

Dissipations, triadic interactions
and mutual coupling terms in $k$ space are shown in
Figure \ref{fig:Flux}.
As expected from the stationary of the flow, 
for each fluid the sum of all terms is close to zero, 
except at the lowest $k$ (for which volume averaging is not performed 
over enough independent realisations due to the limited time scale of
integration). 
In the inertial range and in the near dissipative range, the triadic 
interaction terms for both fluids are the same at first order, 
as expected from $\vn \simeq \vs$. 
In the normal fluid, the triadic term is roughly compensated by the 
viscous dissipation. In the superfluid, the viscous dissipation 
is $\nu_{n}/\nu_{s}=100$ times smaller, and the triadic term is 
compensated instead by the coupling term. 
For consistency, we check that this coupling has a second order effect 
on the normal flux budget: we find 
$\vert M_{n \to s} / M_{s \to n} \vert \simeq \rho_{s}/\rho_{n}=0.1$. 
We conclude that in the inertial range
at high temperatures: the slip velocity 
allows energy to leak from the superfluid 
with a flux which mimics the normal fluid viscous dissipation, 
so that both fluid end up with similar behaviour. This process remains compatible with a strong locking of the 2 fluids in the limit of high Reynolds numbers.

\subsection{The dissipation cut--off in the two--fluids cascade}

In ordinary Navier-Stokes homogeneous turbulence, the viscous
cut--off scale 
$\eta$ is determined by the energy injection $\epsilon$
at large scale $d_0$ and by the kinematic viscosity $\nu$:

\begin{equation}
\label{eq-eta}
\eta^4 \approx \frac{\nu ^3}{\epsilon}. 
\label{eq:cutoff}
\end{equation}

With the exception of the ``quasi-Euler'' case, the spectra 
of $\vn$ and $\vs$ are computed with 
the same viscosities $\nu_{n}$ and $\nu_{s}$ and the same total energy 
injected per unit volume at wavenumber $k\simeq 2$. The collapse of all spectra 
onto the same curve at low $k$ shown in Fig. \ref{fig:Spectre}b 
indicates that the energy injected is 
efficiently redistributed between the two fluids, and that both fluids 
have the same integral scale $d_0$ (as evident in Fig.\ref{fig:Visu}). 
Contrary to what happens in single--fluid Navier-Stokes turbulence, 
Fig.~\ref{fig:Spectre}b shows that the superfluid and normal fluid
cutoff--scales are temperature dependent: it is apparent that
the lower is $T$, the more extended is 
the inertial range cascade. Therefore, if one defines the Reynolds number
using the classical relation with the ratio of
large and small scales of the cascade, $Re=(\eta/d_0)^{-4/3}$, 
one finds that $Re$ is temperature dependent.
%\footnote{As the temperature is lowered, the ratio $\rho_{s} / \rho_{n}$ 
%increases. Therefore, the normal fluid molecular viscosity $\rho_{n} \nu_{n}$
%decreases and one can wonder if this decrease could account for the 
%extension of the cascade. This seems unlikely because the superfluid 
%molecular viscosity $\rho_{s} \nu_{s}$ increases accordingly 
%and the relative fraction of superfluid is larger and larger. 
%This second effect is therefore expected to be the dominant one.}. 
Evidently, ordinary turbulent relations such as Eq.\ref{eq-eta} are 
not valid in our two--fluids system.

We now argue that the temperature dependence of $\eta$ and $Re$ observed 
for finite $\nu_{s}$ will be present in the limit $\nu_{s}=0$, and 
 will reflect the temperature-dependent efficiency of the energy 
transfer from the superfluid to the normal by mutual friction. 
This process is independent of $\nu_{n}$, and more generally 
is independent of the dissipation mechanism in the normal fluid. 
It can be accounted by a temperature-dependent effective superfluid 
viscosity $\nu_{eff}$.

As a starting point, we note in Fig.~\ref{fig:Spectre}b
that at small enough scales $\vert \vn \vert << \vert \vs \vert$.
At these small scales the mutual friction force in the equation for $\vs$ 
simplifies to

\begin{equation}
-\frac{\rho_n}{\rho}\fns
=-\frac{\rho_n}{\rho} \frac{B}{2}  \vert \bom \vert (\vn-\vs)
\approx \alpha  \vert \bom \vert \vs,
\end{equation}

\noindent
where $\alpha=\rho_n B/(2 \rho)$\cite{Schwarz1988}.
In our model, the prefactor $\vert \bom \vert$ in the mutual 
friction accounts for the local absolute vorticity of the vortex 
tangle, which, following Ref.~\cite{Vinen2002}, can be related to
the vortex line density by $\vert \bom \vert =\kappa L$. 
If the vortex lines are smooth on length scales 
of order $\ell \approx L^{-1/2}$,
the quantity $\ell$ measures the inter-vortex spacing 
and corresponds to the cut-off scale of superfluid turbulence. We conclude
that at sufficiently small scales

\begin{equation}
\label{eq:viscoeff}
-\frac{\rho_n}{\rho}\fns 
\approx \alpha \kappa L \vs
\approx \nu_{eff} ( \frac{\vs}{\ell^2}).
\end{equation}

\noindent 
where the analogy between ${ \vs}/{\ell^2}$ and the magnitude of 
$\nabla^2 \vs$ has suggested the introduction of an effective
superfluid viscosity:

\begin{equation}
\label{eq:viscoeffExpression}
\nu_{eff} = \alpha \kappa,
\end{equation}

\noindent
This effective viscosity is relevant at scales smaller or equal to
$l$, at which superfluid energy is transferred into the
normal fluid by mutual friction.

When comparing our DNS model to experiments we must remember that
in our continuous model the identification $L=\vert \bom \vert/\kappa$
neglects random vortex filaments and Kelvin waves of wavelength shorter
than $\ell$ (which become particularly important at very low temperatures),
thus underestimating the vortex line density. It is therefore more  proper 
to relate $\ell$ not to the total $L^{-1/2}$ but to
$L_{\parallel}^{-1/2}$, defined in our previous paper \cite{Roche2008}
as the vortex line density associated with the polarised part of
the vortex tangle. Eq.~\ref{eq:viscoeffExpression} thus becomes

\begin{equation}
\label{eq:viscoeffExpressionGeneralized}
\nu_{eff} = \frac{L}{L_{\parallel}} \alpha \kappa
\label{eq:complete}
\end{equation}

\noindent
where the ratio $L/L_{\parallel} \geqslant 1 $ is expected to be of order one
at high temperature and possibly significantly larger than one
in the limit $T \to 0$. This ratio measures the wiggleness 
of the tangle at small scale, and can be interpreted as a measure
of its fractal dimension\cite{fractal}. 

\begin{figure}[!ht]
\includegraphics[width=\linewidth]{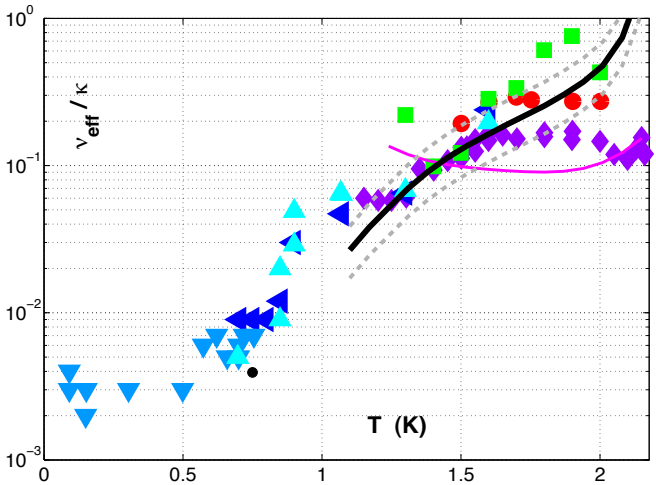}
\caption{[Colours online]. 
Effective kinematic viscosity in units of $\kappa$. 
Isolated symbols: Superfluid effective viscosity defined by 
$\nu_{eff}^{\prime}=\epsilon / (\kappa L)^2$ and measured in 
decaying turbulence experiments; [blue] diamonds \cite{Stalp2002}, 
[red] circles \cite{Niemela:JLTP2005}, 
[green] squares  \cite{Chagovets:PRE2007} and 
triangles  \cite{Walmsley:PRL2007,Walmsley:PRL2008}. 
Lines: analytical models; 
thick [black] line: present model (Eq.\ref{eq:complete}) 
with prefactor $L/L_{\parallel}=2 $ ( isolated bullet [black] for $L/L_{\parallel}=30$),
thin [purple] continuous line: $\mu / ( \rho \kappa)$ 
where $\mu$ is the dynamic viscosity of the normal fluid; 
dashed [grey] lines: microscopic model Eq.~66 from Ref.~\cite{Vinen2002} 
computed with prefactor $R_{0}/a_{0}=10^6$ and $10^4$}
\label{fig:ViscoEffective}   
\end{figure}

An alternative derivation of an effective superfluid
viscosity has been proposed in Ref.~\cite{Vinen2002} (page 216) from 
the more microscopic point of view of the friction of individual vortex lines. 
Fig.\ref{fig:ViscoEffective} shows that our macroscopic model of
$\nu_{eff}$ (thick black line) is in excellent agreement with
the microscopic model of Ref.~\cite{Vinen2002}(thin grey dashed lines).
In the figure
our model is plotted with a wiggleness prefactor $L/L_{\parallel}=2$,
and the model of Ref.~\cite{Vinen2002} contains
a logarithmic prefactor which has been estimated for two different 
sets of parameters for laboratory turbulent flows.
%the excellent agreement with the 
%more refined microscopic model suggests that our model captures reasonably 
%well the underlying physical processes, which justifies a-posteriori 
%the main hypotheses : averaging out $\vn$ in the friction 
%force at small enough scales.

Another (experimental) definition of effective superfluid viscosity 
$\nu_{eff}^{\prime}=\epsilon / (\kappa L)^2$ has been proposed
in recent studies of decaying turbulence experiments \cite{Stalp2002, 
Niemela:JLTP2005, Chagovets:PRE2007,Walmsley:PRL2007,Walmsley:PRL2008}.
These values of $\nu_{eff}^{\prime}$ are shown in
Fig.\ref{fig:ViscoEffective} as symbols. The agreement between
$\nu_{eff}$ and $\nu_{eff}^{\prime}$ is good. Since $B$ is 
approximately constant with temperature above 1K, the strong
temperature dependent of the effective viscosity arises from  the
temperature dependence of $\rho_n/\rho$, which suggests that the ratio
$L/L_{\parallel}$ has little temperature dependence above 1K.

Although what happens below $1~K$ goes beyond the scope of this study,
we remark that there is a qualitative good agreement between 
experimental data below 1K and the low temperature extrapolation of 
the predicted effective viscosity. A wiggliness prefactor $L/L_{\parallel} \simeq 30$ would allow to account for the measured effective viscosity down to 750 mK typically. 
This suggests that 
the cross-over between zero-temperature and finite temperature quantum 
turbulence occurs at a lower temperature than the usual 
estimation of $1~\rm K$ based on phonon (normal fluid) mean free path 
considerations. In other words, it suggests that a relative small
concentration of normal fluid (significantly lower than 1 percent) 
still produces a dissipation which is
comparable to other effects (Kelvin waves cascade, vortex reconnections)
which are specific of the superfluid.
Although our continuous model is no longer justified 
for the normal fluid below $1~\rm K$, this observation is not inconsistent 
with it, because $\nu_{eff}$ is derived regardless 
of the normal fluid dynamics. 

Finally, we speculate on the form of the classical Eq.\ref{eq-eta} 
for superfluids. 
Following the analogy between the mutual friction at 
small scale and the viscous dissipation (Eq.\ref{eq:viscoeff}), 
the kinetic energy which is ``removed'' from the superfluid
at small scales is
\begin{equation}
\label{eq:epsi}
\epsilon  \approx \nu_{eff} ( \frac{ v_{s}^2}{\ell^2}).
\end{equation}

Substituting $v_{s} \approx \kappa / \ell$. 
into Eq.\ref{eq:epsi}, we recognise 
the alternative definition $\nu_{eff}^{\prime}$. 
Combining Eq.\ref{eq:epsi} and Eq.\ref{eq:viscoeffExpressionGeneralized}, 
we obtain the following superfluid counterpart of the classical 
Eq.\ref{eq-eta}:

\begin{equation}
\ell^4 = ( \frac{\rho_n}{\rho} \frac{B}{2} ) 
\frac{\kappa ^3}{\epsilon} \frac{L}{L_{\parallel}}
=\frac{\alpha \kappa^3}{\epsilon} \frac{L}{L_{\parallel}},
\label{eq:superfluid-cutoff}
\end{equation}

\noindent
where the wiggleness parameter $L/L_{\parallel}$ is of order one for 
$T>1$ K and possibly larger at lower temperatures.
Note that Eq.~\ref{eq:superfluid-cutoff} contains an 
implicit temperature dependence through $B$ and ${\rho_{n}}/{\rho}$ 
(and possibly through $L/L_{\parallel}$ in the low temperature limit).

A quantitative comparison of Eq.~\ref{eq:superfluid-cutoff}
with our numerical simulations is impossible due to the finiteness 
of $\nu_{s}$ in particular, but we find a good 
qualitative agreement. 
Eq.~\ref{eq:superfluid-cutoff} allows us to
predict the temperature dependence of the depth of the
turbulent cascade, and to define a
"superfluid Reynolds number" using the 
separation of large and small scales.

\section{Conclusion and Perspectives}

We have introduced a new two--fluids DNS model to study quantum
turbulence in a self--consistent way. The model is based on the
continuum approximation, and we have discussed its advantages
and limitations.  Our numerical results 
support the current understanding \cite{Vinen2002} that in quantum
turbulence the superfluid and the normal fluid are strongly coupled.
In addition, our results at high temperature show the slip velocity 
peaks in the
dissipation scales, and that whereas in the normal fluid the triadic
interaction is balanced by the viscous dissipation, in the superfluid
it is balanced by the mutual friction. We have also found that
the usual turbulence relation (Eq.\ref{eq:cutoff}) which relates the
cutoff scale to the energy injection at large scale $\epsilon$ and the
kinematic viscosity $\nu$ is not valid in our two--fluids system.
Finally, we have found that the energy transfer from the 
superfluid to the normal fluid
does not depend on the normal fluid's dissipation, but it can be accounted
by a temperature--dependent effective superfluid viscosity $\nu_{eff}$,
which we have calculated and which is in good agreement with other
estimates.  Using this quantity, we have obtained the superfluid
equivalent (Eq.~\ref{eq:superfluid-cutoff})
to the classical formula (Eq.~\ref{eq:cutoff})
which relates the energy injection to the dissipation scale. 

In discussing our result we have introduced the wiggleness parameter
$L/L_{\parallel}$  which is or order one above $1~\rm K$ but may
be larger at smaller temperatures. We speculate that $L/L_{\parallel}$
is related to the fractal dimension of the tangle; its increase in the low
temperature limit may explain the saturation of $\nu_{eff}$ for
$T \to 0$. Further work will solve this issue.

In two recent experiments \cite{RocheEPL2007, Bradley:PRL2008}, 
the spectrum of vortex line density in turbulent superfluids was 
found to differ from its classical counterpart: the spectrum of 
local enstrophy. A proposed interpretation assumes that only the 
``polarised'' contribution of the vortex tangle mimics the 
classical enstrophy \cite{Roche2008}. The present numerical model, 
which only accounts for the polarised contribution, gives results 
consistent with this picture :  the spectrum of the superfluid vorticity $| \bom |$ (Fig.~\ref{fig:Spectre}a)
is indeed similar to corresponding spectra in classical fluids. 

Future work will also attempt to test further the proposed 
interpretation\cite{Roche2008} by adding to the current DNS
two--fluids model an equation for a scalar field field accounting for 
the density of the unpolarised contribution of the vortex tangle.
Another important problem to address is the decay of turbulence, which is
receiving much experimental attention.

\acknowledgments 
We thank L. Chevillard, P. Diribarne and J. Salort for their inputs. 
Computations have been performed by using the local computing facilities at ENS-Lyon (PSMN) and at the french national computing center CINES.%(contract c2009025088 ??XXX)
This work was made possible with support from the EPSRC (EP/D040892) and the ANR (TSF and SHREK).

\end{document}